\documentclass[conference]{IEEEtran}
\IEEEoverridecommandlockouts
\usepackage{amsmath,amssymb,amsfonts}
\usepackage{algorithmic}
\usepackage{graphicx}
\usepackage{textcomp}
\usepackage{xcolor}
\usepackage{bm}
\usepackage{subfigure}
\usepackage{hyperref}

\def\BibTeX{{\rm B\kern-.05em{\sc i\kern-.025em b}\kern-.08em
    T\kern-.1667em\lower.7ex\hbox{E}\kern-.125emX}}
\begin{document}

\title{Dynamic Changes of Brain Network during Epileptic Seizure
}

\author{\IEEEauthorblockN{Atefeh Khoshkhahtinat} \IEEEauthorblockA{\textit{Department of Electrical Engineering   }\\
\textit{Sharif University of Technology}\\
Tehran, Iran \\
 \href{mailto:atefe.khoshtinat@yahoo.com}{atefe.khoshtinat@yahoo.com}}
 \and
 \IEEEauthorblockN{Hoda Mohammadzadeh}
 \IEEEauthorblockA{\textit{Department of Electrical Engineering }\\
 \textit{Sharif University of Technology}\\
Tehran, Iran \\
 \href{mailto:hoda@sharif.edu}{hoda@sharif.edu}}

\IEEEauthorblockA{\textit{}}
}

\maketitle

\begin{abstract}

Epilepsy is a neurological disorder identified by sudden and recurrent seizures, which are believed to be accompanied by distinct changes in brain dynamics. Exploring the dynamic changes of brain network states during seizures can pave the way for improving the diagnosis and treatment of patients with epilepsy. In this paper, the connectivity brain network is constructed using the phase lag index (PLI) measurement within five frequency bands, and graph-theoretic techniques are employed to extract topological features from the brain network. Subsequently, an unsupervised clustering approach is used to examine the state transitions of the brain network during seizures. Our findings demonstrate that the level of brain synchrony during the seizure period is higher than  the pre-seizure and post-seizure periods in the theta, alpha, and beta bands, while it decreases in the gamma bands. These changes in synchronization also lead to alterations in the topological features of functional brain networks during seizures. Additionally, our results suggest that the dynamics of the brain during seizures are more complex than the traditional three-state model (pre-seizure, seizure, and post-seizure) and the brain network state exhibits a slower rate of change during the seizure period compared to the pre-seizure and post-seizure periods.

\end{abstract}


\section{\textbf{Introduction}}

%Epilepsy is a highly prevalent neurological disorder characterized by recurrent and sudden seizures, affecting roughly 70 million people around the world. Tracking the evolution of seizures can provide valuable insights into the mechanisms underlying seizure generation, propagation, and termination. Additionally, it has the potential to enhance the treatment of patients with epilepsy.

%Seizures are commonly conceptualized as a hypersynchronous state of neural activity throughout the brain, driven by strong functional connections within the epileptic network. Extensive research has been dedicated to studying seizure dynamics and investigating changes in synchronization among cortical areas, aiming to enhance our understanding of seizure generation. The focus of this research is to unravel the complex mechanisms underlying seizures By exploring the dynamics of synchronization and its alterations during seizures, researchers strive to advance our knowledge of epilepsy and develop more effective strategies for seizure management and treatment.

Epilepsy is a highly prevalent neurological disorder that affects approximately 70 million individuals worldwide.  According to the World Health Organization (WHO), it is estimated that around 40\% of the patients suffer from drug-resistant epilepsy (DRE) \cite{kwan2011drug}. Epilepsy is defined by recurrent and sudden seizures, accompanied by unusual behavior, sensations and sometimes loss of consciousness which can significantly impact the quality of epilepsy patient life \cite{cavanna2009brain}. Monitoring the evolution of seizures can provide valuable insights into the underlying mechanisms of seizure generation and propagation. Moreover, such monitoring holds the potential to improve the treatment of patients with epilepsy.

Seizures have long been conceptualized as a hypersynchronous state of neural activity throughout the brain, resulting from strong functional connections within the epileptic network \cite{jiruska2013synchronization}. This hypersynchrony is believed to contribute to the initiation and spread of seizures.  Recognizing the importance of synchronization in seizure dynamics, extensive research efforts have been dedicated to studying seizure dynamics and investigating changes in synchronization among cortical areas. By examining the dynamics of synchronization and exploring its alterations during seizures, researchers aim to unravel the complex mechanisms that drive epileptic activity.

Various methods have been employed to quantify the synchronization between different brain regions and establish functional brain networks. These methods encompass coherency \cite{nunez1997eeg}, mutual information \cite{shoushtari2021emotion}, partial-directed coherence \cite{li2016localization}, directed transfer function \cite{kim2014combined}, and phase synchronization \cite{mormann2003epileptic}. In the context of epileptic brain activities, which exhibit nonlinear characteristics, phase synchronization stands out as a well-suited metric for quantifying the synchronization between various brain areas \cite{hurtado2004statistical}. This preference stems from the ability of phase synchronization to capture the complex and nonlinear dynamics observed in epileptic seizures.

Investigating brain networks in epilepsy has gained significant attention, with the application of graph theory techniques for characterizing these networks \cite{khambhati2015dynamic,6675620,burns2014network,liu2018phase}.  In the field of neuroscience, the brain is represented as a graph, where nodes represent different brain regions and edges represent the connections between them \cite{rubinov2010complex}.  By utilizing graph theory \cite{reijneveld2007application}, various topological features of the network can be extracted, including path length, global efficiency, clustering coefficient, and modularity \cite{ponten2007small}. These measures enable the assessment of functional integration, segregation, and centrality of individual brain regions. Analyzing these features facilitates the examination of how the network topology changes throughout different stages of epilepsy, such as interictal, preictal, ictal, and postictal periods.

\textbf{Contribution of This Work.} In this paper, our objective is to examine the dynamics of brain network states during the occurrence of seizure using recorded EEG signals from patients. To achieve this, we utilize the Phase Lag Index (PLI) to construct the brain connectivity network and extract its topological features using graph analysis methods. Subsequently, we employ unsupervised clustering technique on these topological features to investigate the reconfiguration of networks across various stages of epileptic seizure. 

The remainder of the paper is organized as follows. In Section II, we provide a comprehensive review of the applications of brain connectivity and graph analysis in the epilepsy domain. Section III presents the dataset used in this study and describes the proposed method in detail. The experimental results are presented in Section IV. Section V concludes the paper.

\begin{figure*}[tp]
    \centering
    \includegraphics[width=1\linewidth]{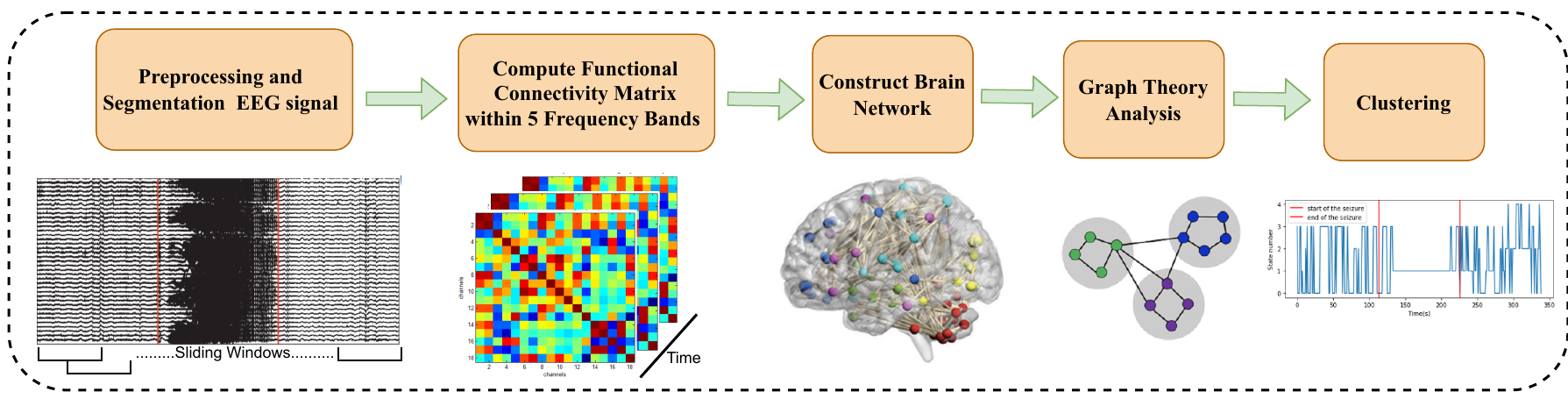}
    \caption{Our proposed framework for analyzing  dynamic changes of brain network states during seizure.}
    \label{fig:network-arch}
\end{figure*}

\section{\textbf{Related Work}}

In the study conducted by Khambhati \emph{et al.} \cite{khambhati2015dynamic}, the researchers hypothesized that each state of the brain could be characterized by a set of feature vectors representing similar patterns of functional connectivity. To explore this hypothesis, they constructed a configuration-similarity matrix to quantify the similarity between the network geometries of each pair of temporal windows. Then, they identified communities within the configuration-similarity matrix using an unsupervised clustering approach. The findings of the study revealed that the functional networks, computed using magnitude-normalized cross-correlation between pairs of sensors, exhibited rapid state transitions during the pre-seizure period compared to the seizure period. Another study \cite{burns2014network} examined the temporal dynamics of brain connectivity before, during, and after seizures using intracranial EEG data. In this work, coherence was used as a measure to estimate functional networks within specific frequency bands, and the eigenvector centrality parameter was calculated to characterize the geometry of the estimated networks. The investigation focused on examining the temporal evolution of eigenvector centrality across the pre-seizure, seizure, and post-seizure periods using a clustering method. The findings of this study suggested that epileptic networks exhibit a limited number of distinct states between which they oscillate.

Liu \emph{et al.} utilized the phase locking value (PLV) measurement to assess the relationship between pairs of ECoG signals and construct a brain network. They assessed  the synchronization dynamics and evolution of network states across the pre-seizure, seizure, and post-seizure periods. Their results demonstrated an increase in synchronization levels prior to seizure termination, which persisted at a high level following seizure termination. Liu \emph{et al.} \cite{liu2017dynamics} examined the transitions between network states of epileptiform discharges in hippocampal slices. By utilizing clustering techniques on degree vectors, they were able to identify distinct network states. Notably, during the ictal-like discharges, they discovered two network states that corresponded to the tonic and clonic phases of epileptic activity. Ma \emph{et al.} \cite{ma2021spatiotemporal} proposed model which investigates the functional interactions among different brain regions and tracks the state transition during the seizure procedure. By leveraging the mutual information (MI) indicator, they construct a dynamic brain network and monitor the changes in seizures by analyzing the parameter characteristics of the resulting networks.

%In the study conducted by Ankit N. Khambhati, it is assumed that each state of the brain could be characterized by a set of feature vectors displaying similar patterns of functional connectivity. To investigate this, they constructed a configuration-similarity matrix to quantify the similarity between the network geometries of each pair of temporal windows and then, communities within the configuration-similarity matrix were identified using an unsupervised clustering approach. The findings indicate that the functional networks, computed by utilizing a magnitude normalized cross-correlation between pairs of sensors, exhibited rapid state transitions during the pre-seizure period compared to seizure period. The [85] explored the temporal dynamics of brain connectivity before, during, and after seizures using intracranial EEG data. They employed coherence as a measure to estimate functional networks within specific frequency bands and calculated the eigenvector centrality parameter to characterize the geometry of the estimated networks. The investigation focused on examining the temporal evolution of eigenvector centrality across the pre-seizure, seizure, and post-seizure periods using an clustering method. The findings of the study suggest that epileptic networks exhibit a limited distinct states that they oscillate between.

In the realm of seizure detection and prediction, analysing of brain network are utilized. In this study \cite{kerr2011multivariate}, stereoelectroencephalography (SEEG) data was analyzed to investigate the characteristics of seizure state. Functional connectivity matrices are generated from the SEEG data by using the coherence measure, and singular value decomposition (SVD) was applied to analyze these matrices. It was observed that the largest eigenvector corresponding to the estimated functional connectivity matrices could potentially serve as an appropriate metric for detecting the onset of seizure activity. Wang \emph{et al.} \cite{wang2022spatiotemporal} introduced a synchronization-based model called the spatiotemporal graph attention network (STGAT) for seizure prediction. By leveraging phase locking value (PLV) to capture spatial and functional connectivity information from EEG channels, the graph representation of the EEG data is constructed. The STGAT model was then employed to learn the temporal correlation properties of EEG sequences and explore the spatial topological information of multiple channels. This approach allows for a dynamic analysis of EEG data, enabling the prediction of seizures. A novel dynamic learning method is introduced \cite{bomela2020real}, which involves inferring a time-varying network from multivariate EEG signals to capture the overall dynamics of the brain network. The topological properties of this network are then quantified using graph theory. The efficacy of this learning method is demonstrated in detecting strong synchronization patterns associated with abnormal neuronal firing during seizure onset. Li \emph{et al.} \cite{li2019transition} investigated the transition of brain function from inte-rictal state to the pre-ictal state preceding a seizure using scalp EEG network analysis. They examined functional connectivity networks derived from EEG signals and implementing clustering techniques across various frequency bands. Their findings revealed that the beta band exhibited the highest clustering performance using the k-medoids algorithm. Moreover, they observed that the pre-ictal state exhibited increased synchronization of EEG network connectivity compared to the inter-ictal state.

This paper \cite{li2022graph} presented a novel graph generative neural network (GGN) model for analyzing scalp EEG signals and
exploring brain functional connectivity. The model generates brain functional connectivity graphs to capture spatialtemporal patterns associated with different types of onset
epilepsy seizures. The main goal of the model is to accurately
classify and identify the specific type of epileptic seizure based
on the extracted features and connectivity patterns

Functional connectivity analysis and graph theory were used to enhance the identification of the epileptogenic zone (EZ) using intracranial EEG data \cite{devisetty2022localizing}. By applying nonlinear correlation and mutual information measures, the study estimated the interaction patterns among EEG signals. The findings indicated that graph properties such as out-degree, out-strength, and betweenness centrality were effective predictors for accurately localizing the epileptogenic region.

In the research conducted by \cite{ponten2009indications}, intracranial EEG signals from eight epileptic patients were analyzed, leading to notable findings concerning network dynamics. The research revealed that during seizures, the network exhibited a higher clustering coefficient and shorter path length, indicating a more organized and structured network. Conversely, during the inter-seizure period, the network appeared random and disordered. These findings suggest a distinct shift in network properties between seizure and non-seizure states, highlighting the dynamic nature of epileptic brain activity.

\section{\textbf{Methods}}
\subsection{\textbf{Dataset}}
The CHB-MIT database \cite{shoeb2009application} used in this study is a publicly available dataset that contains scalp EEG signals obtained from 23 pediatric patients diagnosed with intractable epilepsy. The dataset includes both male and female participants, with ages ranging from 3 to 22 for males and 1.5 to 19 for females. Seizure onsets in the database are categorized into three types: focal seizures, lateral seizures, and generalized seizures. The EEG signals were recorded at a sampling rate of 256 Hz and with a resolution of 16 bits. The recordings were performed using a varying number of channels, with most cases using 23 channels, although in some instances 24 or 21 channels were used. The placement of electrodes or channels followed the International 10-20 system, ensuring consistency across recordings.

\begin{table}[]
\caption{ Detailed clinical information for the 7 patients.}

\scalebox{1.2}{
\begin{tabular}{c c c c} % centered columns (4 columns)
\hline\hline %inserts double horizontal lines
Patient  & Gender & Age &  Number of Seizures \\ [0.2ex] % inserts table
%heading
\hline % inserts single horizontal line
Chb01 & Female & 11 & 7 \\ % inserting body of the table
Chb02 & Male & 11 & 3 \\
Chb03 & Female & 14 & 7 \\
Chb05 & Female & 7 & 5 \\
Chb07 & Female & 14.5 & 3 \\
Chb08 & Male & 3.5 & 3 \\
Chb10 & Male & 3 & 7 \\  % [1ex] adds vertical space
\hline %inserts single line
\end{tabular}}
\label{result}
\centering
\end{table}

\subsection{\textbf{Preprocessing}}
The recorded raw data is preprocessed using Makoto's pipeline \cite{miyakoshi2018makoto}, which consists of four steps: 1) Applying a high-pass filter with a cutoff frequency of 0.5 Hz to remove baseline drifts in the signal. 2) Changing the reference system of the data from a bipolar system to a common average reference system. 3) Eliminating power line noise at 60 Hz using the CleanLine algorithm \cite{bigdely2015prep}. 4) Employing the Independent Component Analysis (ICA) method to remove any remaining artifacts. 

%These steps ensure the data is prepared for further analysis and improve the quality of the EEG signals.

A total of seven patients' data is examined in this study. The reason for not utilizing data from all patients is the lack of specified channel names for some individuals. Due to the unavailability of channel names, it is not possible to re-reference the signals from a bipolar system to a common average reference system. Detailed information about the patients can be found in Table I.

To investigate dynamic changes in brain connectivity during seizures, after preprocessing the raw data, we select the same time intervals of the seizure period before seizure onset and after seizure termination. These intervals are referred to as the pre-ictal and post-ictal periods, respectively. Then, the EEG signals are filtered using finite impulse response (FIR) filters obtained from EEGLAB \cite{delorme2004eeglab} to extract specific frequency bands, including theta (4-8 Hz), alpha (8-13 Hz), beta (13-30 Hz), low gamma (30-60 Hz), and high gamma (60-80) bands. The utilization of these filters ensures minimal phase distortion enabling accurate analysis of phase synchronization in the selected frequency bands.

\subsection{\textbf{Functional Connectivity Measure}}

Phase Lag Index (PLI) is a phase synchronization index that is commonly used to analyze the functional connectivity of the brain \cite{stam2007phase}. In the domain of studying brain connectivity, coherence and phase locking value (PLV) are often employed as measures to to quantify the degree of the synchronization between different brain regions. However, these indices can be misleadingly increased due to a phenomenon known as volume conduction. To overcome this issue, PLI is introduced as an alternative phase synchronization index. 
To compute the PLI between two signals, the first step involves obtaining the Hilbert transform of each signal. The Hilbert transform is a mathematical operation that provides the instantaneous phase information of the signals at each time point. The Hilbert transform is calculated as follows:
\begin{equation}
H[x(t)]= \frac{1}{\pi} P.V. \int_{-\infty}^\infty \frac{x(\tau)}{t-\tau} d\tau, 
\end{equation}

where $P.V.$ corresponds to the Cauchy principal value \cite{hardy1909theory}. Then, the instantaneous phase $\phi(t)$ can be computed by:

\begin{equation}
\phi(t)= arctang \frac{H[x(t)]}{x(t)}. 
\end{equation}

 When the phase of each signal is calculated, the PLI between a pair of signals $x_1(t)$ and $x_2(t)$ can be estimated as :

\begin{equation}
\text{PLI}= \frac{1}{N}|\sum_{t=1}^{N} sign({\phi}_1(t)-{\phi}_2(t))|, 
\end{equation}

where ${\phi}_1(t)$ and ${\phi}_2(t)$ represent the instantaneous phases of signals $x_1(t)$ and $x_2(t)$, respectively, and $N$ is the total number of samples in each time window. The PLI value ranges between $0$ and $1$, where $0$ indicates no synchronization between the signals and $1$ indicates maximal synchronization.

\subsection{\textbf{Graph Theory}}

Graph theory is employed to assess specific brain network properties. A weighted undirected graph associated with a set of EEG channels can be denoted by the triple $G=(\nu,\varepsilon, A)$. In this representation, $\nu$ refers to the set of EEG channels  which serve as the vertices in the graph, $\varepsilon$ is the set of edges which represents the connections between the channels, and $A$ is the weighted adjacency matrix which corresponds to the connectivity between the EEG channels.

The topological characteristics of a graph can be categorized into two main groups: functional segregation and integration. Functional segregation in the brain refers to the capacity for specialized processing to happen within closely interconnected groups of brain regions. Measures of segregation are typically computed based on the number of triangles in the network, with a higher number of triangles indicating a higher level of segregation. On the other hand, functional integration in the brain relates to the ability to effectively combine and integrate information from different brain regions \cite{rubinov2010complex}. Measures of integration often utilize the concept of path length to assess the efficiency of information flow between brain regions. In the following, we will explain the metrics associated with these two groups.

\subsubsection{\textbf{Characteristic Path Length}}
The characteristic path length was used as a measure to quantify the efficiency of information transfer between different regions of the brain \cite{rubinov2010complex}. The characteristic path length is defined as:

\begin{equation}
L= \frac{1}{N(N-1)} \sum_{i=1}^{N} {\sum_{j \neq i}^{N}} l_{ij} , 
\end{equation}
where $l_{ij}$ is the weighted shortest path between nodes $i$ and $j$, and $N$ is the total number of nodes.

\subsubsection{\textbf{Clustering Coefficient}}
The clustering coefficient is a measure that quantifies the tendency of nodes (brain regions) to form densely interconnected clusters \cite{stam2009graph}. It is a fundamental measure for assessing the degree of segregation and the potential for information processing and communication within a brain network. In a symmetric weighted network, the clustering coefficient for each node is defined as: 
\begin{equation}
C_i= \frac{\sum_{k \neq i}^{N} {\sum_{{l \neq i},{l \neq k}}^{N}} w_{ik} w_{il} w_{kl}}{\sum_{k \neq i}^{N} {\sum_{{l \neq i},{l \neq k}}^{N}} w_{ik} w_{il}}, 
\end{equation}

%The clustering coefficient quantifies the tendency of nodes (brain regions) to form densely interconnected clusters.  It is a basis measure to quantify the degree of segregation of brain regions, as well as the potential for information processing and communication within the brain network. The  clustering coefficient for each node in symmetric weighted network is defined as:

where $C_i$ is the clustering coefficient of node $i$, and $w_{ij}$ is the weight between nodes $i$, $j$. The clustering coefficient ranges from zero to one. A high clustering coefficient indicates a strong interconnectivity among neighboring brain regions. In contrast, a low clustering coefficient suggests a more random pattern of connectivity between adjacent brain regions.

\subsection{\textbf{Clustring Method}}

To examine the transitions of brain states, an agglomerative hierarchical clustering method is employed on the feature vectors over time, utilizing Ward's criterion \cite{ward1963hierarchical}. In this clustering approach, each data point is initially treated as an individual cluster. During an iterative process, the two similar clusters are merged until only a single cluster or a predetermined number of clusters is achieved. The objective function used to determine the optimal number of clusters is defined as follows:

\begin{equation}
Loss= \sum_{i=1}^{k} \sum_{x \in C_{i}} (x-\mu_{i})^2, 
\end{equation}

where $k$ denotes the number of clusters, ${\mu}_{i}$ is the centroid of cluster $i$, and $C_i$ represents cluster $i$. The defined objective function is to minimize the sum of squared distances between data points $x$ and their respective cluster centroids ${\mu}_i$. The optimal number of clusters is determined by identifying the knee point in the loss function curve, plotted against the number of clusters, where the second derivative is the largest.

\subsection{\textbf{Statistical Analysis}}
The Wilcoxon rank-sum test \cite{wilcoxon1992individual} is used to compare all pairs of groups. A significance level of $p < 0.05$ is applied to determine whether there are statistically significant differences between the compared groups.

\section{\textbf{Experimetal Results}}
%To explore brain network transition of each patient within each frequency band, 2/3 number of each patient' seizures are chosen as a training set and remaining seizures are chosen as a test set. In the training phase, a set of clusters are obtained by applying agglomerative hierarchical clustering to the set of extracted feature vecors, including characteristic path and clustering coefficent for each 1s window. During the test phase, the feature vectors of test seizure are compared to the learned cluster centroids to classify them into obtained categories. Figure shows how  clustering the feature vectors of gamma band for patiant 34. As shown in , the optimal number of clusters are determined by using the second derivative of SSE cost function. Then, the feature vectors of test seizure are clusters using obtained centroids. According to Figure, the optimal number of clusters is determined to be 5, which suggests that the dynamics of the brain are more complex than the conventional model with three seizure states: pre-seizure, seizure, and post-seizure. The network states switched among state 1, state 2, state 3, state 4, and state 5 frequently during the preseizure and postseizure periods. However, the frequency of network state changes decreased significantly during the seizure period, and the state 4 dominated the most of seizure periods.

\begin{figure}
    
  \centering
  \scalebox{.7}{\includegraphics{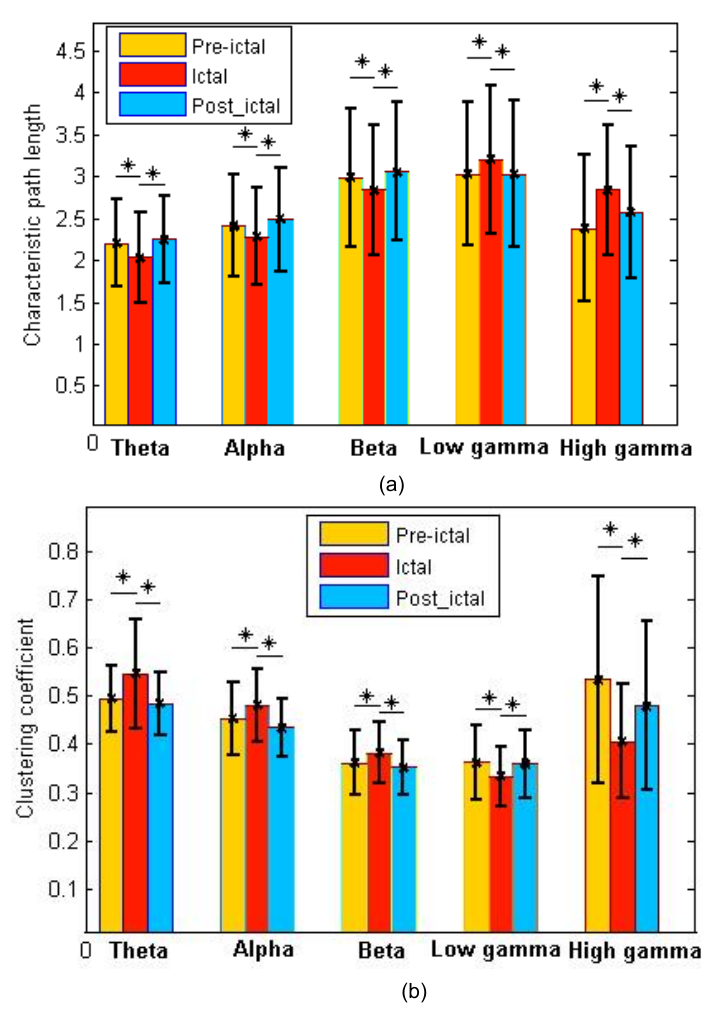}}
  \caption{Statistical analysis results of topological features within five different frequency bands (theta, alpha , beta, low gamma and high gamma). * denotes ${P}< 0.05*$.}
  \label{fig:qplot}
\end{figure}

For each patient, the functional connectivity matrix based on the PLI measure is computed within five different frequency bands (i.e. theta, alpha, beta, low gamma and high gamma) on 2-s sliding windows with  50\% overlap. The weighted undirected graph is then constructed for each 1-s window using the connectivity matrix, where the nodes represent EEG channels and the edges are defined by the PLI values. In this study, the characteristic path length and clustering coefficient metrics are exploited to extract the topological features of the brain graph.
\subsection{\textbf{Investigating the functional connectivity network changes of the brain during seizure}} We compare the brain network properties of the seizure (ictal) period with both the pre-ictal and post-ictal periods in different frequency bands by using the total seizure data from 7 patients. As shown in Figure 2, in the theta, alpha, and beta bands, the ictal period shows significantly shorter characteristic path lengths and higher clustering coefficient compared to the pre-ictal and post-ictal periods. However, in the gamma bands, the opposite trend is observed, with the ictal period exhibiting lower clustering coefficients and higher characteristic path length compared to the other periods.

The increase in the clustering coefficient in the theta, alpha, and beta bands during the seizure period suggests a higher level of correlation or synchronization among different brain regions and an increase in local connectivity. Conversely, the decrease of this measure in the gamma bands indicates a reduction in local connectivity during seizure activity. This reduction in local connectivity is attributed to the anti-binding role of seizure in the gamma bands, particularly the spike-wave components of the seizure signal. The excessive temporal binding ("over-binding") exists between neuronal populations in the gamma bands during the pre-ictal period which triggers epileptic seizures as a protective mechanism to prevent severe consequences of excessive synchrony \cite{medvedev2001temporal}. The reduction of characteristic path length  in the theta, alpha, and beta bands during the seizure period suggests that information is being exchanged through shorter paths in the brain during the seizure interval, allowing for rapid integration. However, in both gamma bands, this parameters increase during the seizure period, indicating a lower level of coherence in the brain connectivity.

\begin{figure}
    
  \centering
  \scalebox{.72}{\includegraphics{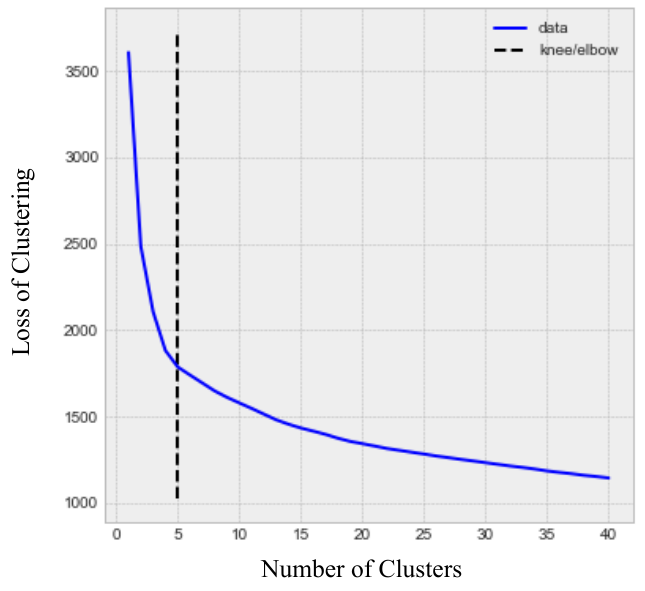}}
  \caption{The clustering loss Function curve with respect to different numbers of clusters. The knee point shows the optimal number of clusters.}
  \label{fig:qplot}
\end{figure}

\begin{figure}
    
  \centering
  \scalebox{.65}{\includegraphics{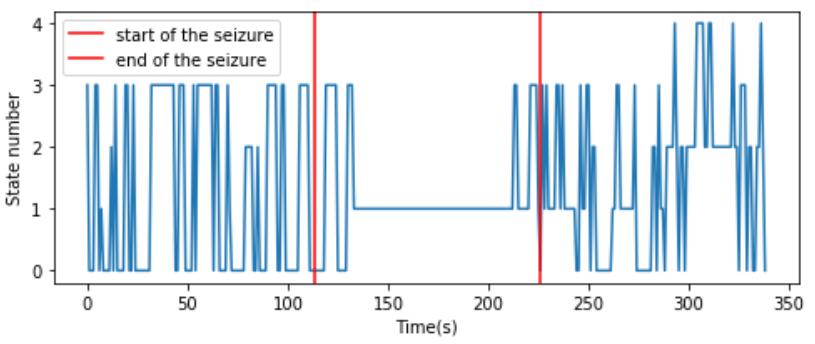}}
  \caption{Transitions of network brain states of patient chb05 during pre-ictal, ictal, and post-ictal periods within high gamma band. }
  \label{fig:qplot}
\end{figure}

\subsection{\textbf{Evaluating the brain network state transition during seizure}}
To investigate the brain network transition for each patient within each frequency band, two-thirds of the patient's seizures are selected as a training set, while the remaining seizures are used as a test set. In the training phase, a set of clusters are obtained by applying agglomerative hierarchical clustering to the set of extracted feature vectors, including concatenation of characteristic path length and clustering coefficients for each 1s window. During the test phase, the feature vectors of the test seizures are compared to the learned cluster centroids to classify them into the obtained categories (comparison is done by computing Euclidean distance). Figures 3, 4, and 5 depict the results obtained for patient chb05, serving as a representative example. Notably, the results from this patient are in line with the findings observed in other subjects. Figure 4 illustrates the clustering of feature vectors in the gamma band for patient chb05. As shown in Figure 3, the optimal number of clusters is determined using the second derivative or knee point of the clustering loss function. The feature vectors of the test seizures are then clustered using the resulted centroids. Figure 3 shows that the optimal number of clusters is determined to be 5, indicating that the dynamics of the brain are more complex than the conventional hypothesis with three states, namely pre-ictal, ictal, and post-ictal. As shown in Figure 4, during the pre-ictal and post-ictal  periods, the brain network undergoes frequent transitions between state 0, state 1, state 2, state 3, and state 4. However, during the seizure period, there is a significant decrease in the frequency of network state changes. In fact, state 1 predominates during most of the ictal period.

Figure 5 illustrates the state transition diagram within the high gamma band during seizure occurrence, which includes pre-ictal, ictal, and post-ictal periods. The diagram reveals five possible states: state 0, state 1, state 2, state 3, and state 4. The arrows between states represent the transition probabilities, denoted as $p_{ij}$. If there is no arrow from state $i$ to state $j$, it indicates that $p_{ij}=0$. As observed, there is a higher probability of remaining in state 1 compared to the other states, suggesting that state 1 represents a stable state. This observation aligns with the findings depicted in Figure 4, where the brain predominantly remains in this stable state throughout the ictal period.

\begin{figure}
    
  \centering
  \scalebox{0.84}{\includegraphics{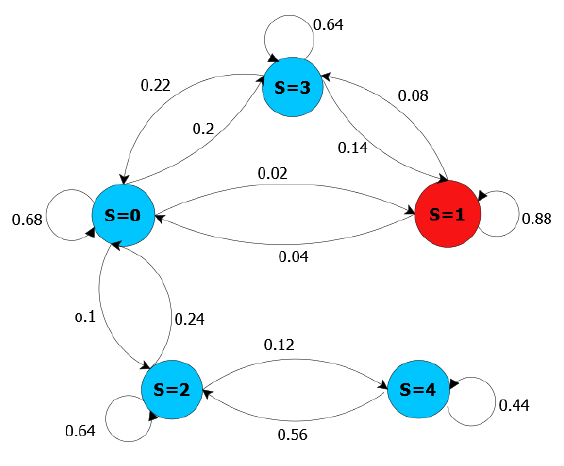}}
  \caption{ A state transition diagram in the high gamma band. }
  \label{fig:qplot}
\end{figure}

\begin{figure}
    
  \centering
  \scalebox{0.97}{\includegraphics{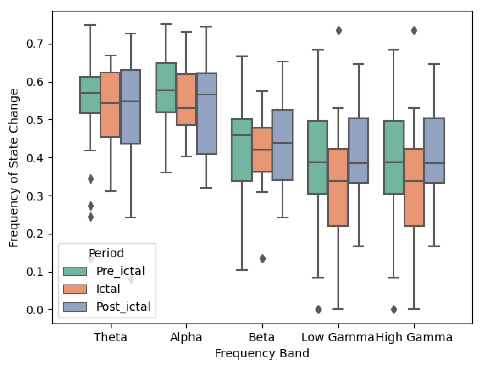}}
  \caption{ The distribution of the network state transition rate in three periods (ictal, pre-ictal, and post-ictal) within five different frequency bands (theta, alpha , beta, low gamma and high gamma). }
  \label{fig:qplot}
\end{figure}

To assess the reconfigurability of the epileptic network in different frequency bands, we calculate the network state change frequency for the pre-ictal, ictal, and post-ictal periods using the results obtained from 35 seizures of 7 patients which results are depicted in Figure 6. By conducting a Wilcoxon rank-sum test, we compare the network transition rate during the ictal period with both the pre-ictal and post-ictal periods. Our results in Table II indicate a statistically significant difference in the rate of change within the gamma frequency bands. As shown in Figure 6,  in these frequency bands, the brain network exhibits a significantly slower rate of change during the ictal period compared to the pre-ictal and post-ictal periods.

\begin{table}[]
\caption{ The Results of Statistical Analysis for network state transition rate. }

\scalebox{0.98}{
\begin{tabular}{c c c } % centered columns (4 columns)
\hline\hline %inserts double horizontal lines
Frequency Band  & P-Value (Pre-Ictal, Ictal) & P-Value (Ictal, Post-Ictal) \\ [0.2ex] % inserts table
%heading
\hline % inserts single horizontal line
Theta & 0.5038 & 0.9170  \\ % inserting body of the table
Alpha & 0.4278 & 0.658  \\
Beta & 0.3481 & 0.8926  \\
Low Gamma & \textbf{0.0081} & \textbf{0.009}  \\
High Gamma & \textbf{0.0042} & \textbf{0.000024}  \\ % [1ex] adds vertical space
\hline %inserts single line
\end{tabular}}
\label{result}
\centering
\end{table}

%For each patient, we tracked the evolution ofthe brain network over consecutive windows (window length: 2.5 s; overlap: 1.5 s)

\section{\textbf{Conclusion}}

In this study, we have investigated the dynamic changes in brain network states during epileptic seizures. By analyzing the functional connectivity brain network using the PLI measurement in multiple frequency bands, we observed significant increases in brain synchronization during the seizure period, particularly in the theta, alpha, and beta bands, while reduction were observed in the gamma bands. These changes in synchronization were associated with alterations in the topological features of functional brain networks. Moreover, our findings suggested that the dynamics of the brain during seizures are more intricate than the traditional three-state model, with a slower rate of change in the brain network state during the seizure period compared to the pre-seizure and post-seizure periods. These analysing contribute to a better understanding of the underlying mechanisms of epileptic seizures and highlight the potential for using dynamic network analysis for improved diagnosis and treatment strategies for patients with epilepsy.

\bibliographystyle{IEEEtran}
\bibliography{sample}
\end{document}